# Multiscale modeling strategy to solve fullerene formation mystery

Andrey M. Popov[*]

*Institute for Spectroscopy of Russian Academy of Sciences, Fizicheskaya Street 5, Troitsk, Moscow 108840, Russia.*

Irina V. Lebedeva

*CIC nanoGUNE BRTA, Avenida de Tolosa 76, San Sebastian 20018, Spain*

*Catalan Institute of Nanoscience and Nanotechnology - ICN2, CSIC and BIST, Campus UAB, Bellaterra 08193, Spain*

Sergey A. Vyrko, Nikolai A. Poklonski

*Belarusian State University, Nezavisimosti Ave. 4, Minsk 220030, Belarus*

---

[*] CONTACT Andrey M. Popov, e-mail: popov-isan@mail.ru, Institute for Spectroscopy, Troitsk, Moscow, Russia

# Multiscale modeling strategy to solve fullerene formation mystery

Since fullerene formation occurs under conditions where direct observation of atomic-scale reactions is not possible, modeling is the only way to reveal atomistic mechanisms which can lead to selection of abundant fullerene isomers (like $C_{60}$-$I_h$). In the present paper we review the results obtained for different atomistic mechanisms by various modeling techniques. Although it seems that atomic-scale processes related to odd fullerenes (such as growth by consecutive insertions of single carbon atoms and rearrangements of the $sp^2$ structure promoted by extra sp atoms) provide the main contribution to selection of abundant isomers, at the moment there is no conclusive evidence in favor of any particular atomistic mechanism. Thus, the following multiscale modeling strategy to solve the mystery of the high yield of abundant fullerene isomers is suggested. On the one hand, sets of reactions between fullerene isomers can be described using theoretical graph techniques. On the other hand, reaction schemes can be revealed by classical molecular dynamics simulations with subsequent refinement of the activation barriers by ab initio calculations. Based on the reaction sets with the reaction probabilities derived in this way, the different atomistic mechanisms of abundant fullerene isomer selection can be compared by kinetic models.

## 1. Introduction

Fullerenes were discovered more than 30 years ago.[1] However, the detailed atomistic mechanism of their formation is still not known. To be more precise, now it is apparent that formation of an *initial arbitrary* fullerene shell occurs via the self-organization process, while atomic-scale processes at the later stage corresponding to selection of *certain abundant* fullerene isomers (like $C_{60}$-$I_h$) remain unclear. Although a set of atomic-scale mechanisms of these processes have been proposed, we do not pursue here the goal to choose and prove one specific mechanism among them. On the contrary, we show that the present knowledge of atomistic mechanisms of fullerene formation is far from the state where the problem is solved and further mammoth job is necessary. We believe that it is hardly probable that atomic-scale mechanisms of selection of abundant fullerene isomers can be revealed by direct observation under experimental conditions (like arc discharge) where this selection take place. Therefore, atomistic modeling is the only way to clarify details of the atomistic mechanisms. However, atomistic modeling alone is not sufficient to unravel the mystery of formation of $C_{60}$-$I_h$ and other abundant fullerene isomers since it would require huge computing resources. The aim of the present paper is to propose the whole strategy to solve this mystery using a multiscale approach which combines various simulation methods such as atomistic modeling, kinetic models and graph techniques.

Fullerene is a molecule with the shape of a closed shell which consists of an even number of $sp^2$ carbon atoms and contains only pentagons and hexagons. Such a structure is a so-called classical fullerene. Here we use the term "odd fullerene" for a molecule with a similar structure but with an extra sp atom instead of one bond (so-called sp defect[2]) or above one bond. Other polygons such as heptagons, octagons and so on are considered as defects of fullerene structure (even fullerenes with such polygons in the $sp^2$ structure are often referred to as nonclassical fullerenes). According to the isolated pentagon rule, abundant fullerene isomers (like $C_{60}$-$I_h$ with icosahedral symmetry and $C_{70}$-$D_{5h}$ isomer) have no adjacent pentagons.[3]

The paper is organized in the following way. In section 2 we consider the formation of initial arbitrary fullerene shells via self-organization. Section 3 is devoted to possible atomistic mechanisms which can lead selection of abundant fullerene isomers starting from an arbitrary fullerene shell and experimental results related with these mechanisms. In section 4 we discuss modeling of fullerene formation. Subsection 4.1 is devoted to inherent features of fullerene formation which lead to problems in atomistic modeling. Different simulation methods used to study fullerene formation and the results obtained are described in Subsections 4.2 and 4.3, respectively. Section 5 is devoted to the proposed strategy to unravel the mystery of formation of abundant fullerene isomers.

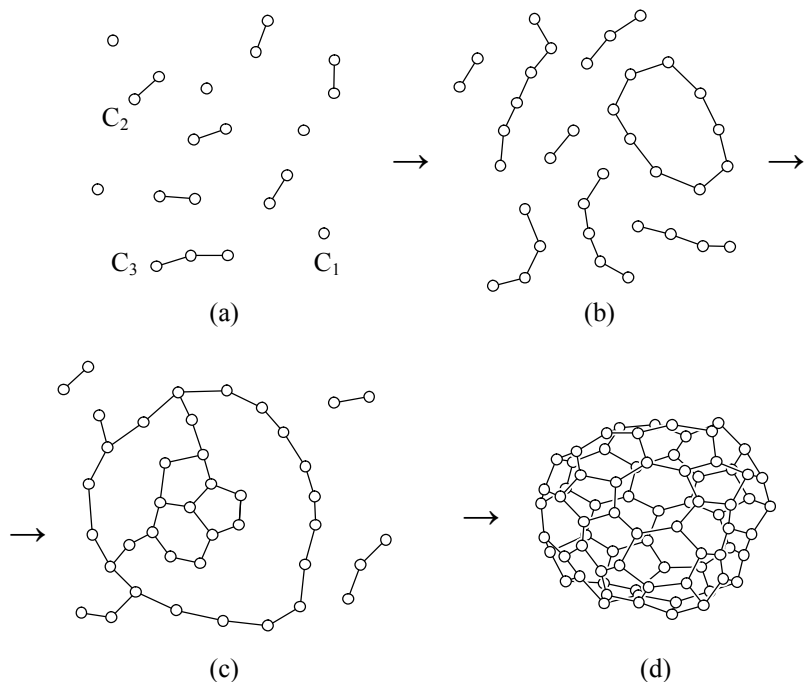

**Figure 1.** Schematic representation of formation of the initial fullerene shell according to the self-organization paradigm: (a) nonequilibrium carbon vapor, (b) formation of carbon chains and a ring, (c) sticking of the chains and ring together into a polycyclic carbon cluster, (d) growth and transformation of the polycyclic carbon cluster into the initial fullerene shell with numerous defects.

## 2. Self-organization paradigm of fullerene formation

Although a set of exotic ways of fullerene formation has been proposed,[4,5,6,7,8] at the moment, the commonly accepted paradigm of fullerene formation is self-organization of a carbon system without precursors of certain structure (see, for example,[9,10] for reviews). Such a paradigm is based on two main statements. First, a fullerene is the ground state of a carbon system consisting of tens or hundreds atoms.[11,12,13] Second, experiments show fullerene formation starting from various initial systems consisting of pure carbon material: fullerenes are formed in hot carbon vapor during arc discharge synthesis[14] or graphite laser ablation[1] as well as under laser ablation of higher carbon oxides in the processes involving at intermediate stages carbon rings $C_{18}$, $C_{24}$, and $C_{30}$ that merge into larger clusters;[15,16] merging of $C_{60}$ fullerenes during laser ablation of the fullerene $C_{60}$ film also leads to formation of fullerenes containing hundreds of atoms;[17] direct transformation of a small graphene flake to a fullerene under action of the electron beam has been revealed;[18] transformation of bicyclic and tricyclic carbon clusters to fullerenes has been observed in the drift tube.[19,20,21,22] Transformation of amorphous carbon clusters to fullerenes has been shown also by molecular dynamics (MD) simulations.[2,23] According to the self-organization paradigm, formation of the initial fullerene shell starting from carbon vapor occurs through the following stages (see Figure 1): first chains grow, then these chains stick together into polycyclic clusters, the polycyclic clusters grow and transform into fullerene shells with numerous defects (see, for example, MD simulations[10,24,25]).

The self-organization paradigm explains only how arbitrary fullerenes are formed. However, the principal mystery related with the fullerene formation is the high yield of certain abundant fullerene isomers. The yield of two of the most abundant fullerene isomers $C_{60}$-$I_h$ and $C_{70}$-$D_{5h}$ in the ratio of approximately 3:1 can achieve up to 26% of carbon soot produced in arc discharge synthesis.[26] Other abundant fullerenes such as $C_{36}$, $C_{50}$, $C_{76}$, $C_{78}$, and $C_{84}$ and so on should be also mentioned (see, for example, Refs. [3,26,27]). The self-organization paradigm does not give any advantage to certain fullerene isomers and even to certain fullerene sizes. On the contrary, the mixture of fullerenes of different sizes with a wide set of different isomers for each fullerene $C_{2n}$ should be formed upon self-organization of any initially chaotic carbon system where fullerenes can be produced. Moreover, the self-organization paradigm does not give any preference to formation of initial fullerene shells consisting of an even number of atoms relative to shells consisting of an odd number of atoms. For example, mass spectra of carbon clusters produced from graphite by laser ablation[27] and using the pulsed microplasma cluster source[28] show that the fraction of clusters with an odd number of atoms is only several times less than the fraction of even clusters. MD simulations of fullerene formation as a result of self-organization of initially chaotic carbon systems also demonstrate that formation of both even and odd fullerenes takes place.[2,24,29,30,31,32]

MD simulations devoted to formation of initial fullerene shells through self-organization of carbon vapor[24,25,29,30,33,34,35,36,37] and transformation of amorphous carbon clusters,[2,23,28] graphene flakes,[8,38] short carbon nanotubes with open ends,[39,40] and small nanodiamond clusters[41] to fullerenes indicate that such initial shells (just after complete formation of the $sp^2$ structure) contain not only adjacent pentagons but also numerous structural defects like heptagons, octagons and so on. There are no reasons to suppose that the initial fullerene shell formed as a result of self-organization has the perfect structure consisting only of pentagons and hexagons, all the more, to suppose absence of adjacent pentagons. Thus, to explain formation of abundant fullerene isomers with certain structure from initial fullerene shells with an arbitrary structure formed as a result of self-organization, it is necessary to investigate atomic-scale reactions of bond rearrangements, insertion and emission of carbon atoms and small molecules which occur after fullerene shell formation.

## 3. Atomistic mechanisms and experimental results for fullerene isomer selection

First we briefly list possible atomistic mechanisms which can contribute into selection of abundant fullerene isomers. We emphasize once more that the present state of fullerene formation studies does not allow us to choose a particular atomistic mechanism, all the more certain atomic-scale reactions which lead to formation of $C_{60}$-$I_h$ and other abundant fullerene isomers. Since $C_{60}$-$I_h$ is the most abundant fullerene isomer, schemes of possible atomic-scale reactions are shown in Figures 2 and 3 by the examples of $C_{60}$-$I_h$ isomer or isomers which can be transformed to $C_{60}$-$I_h$ isomer by one simple reaction. These examples, however, do not mean that atomic-scale reactions that actually take place in real systems are known in detail. General schemes of the reactions without specification of the numbers of atoms in polygons of $sp^2$ fullerene structure are also shown in Figures 2 and 3.

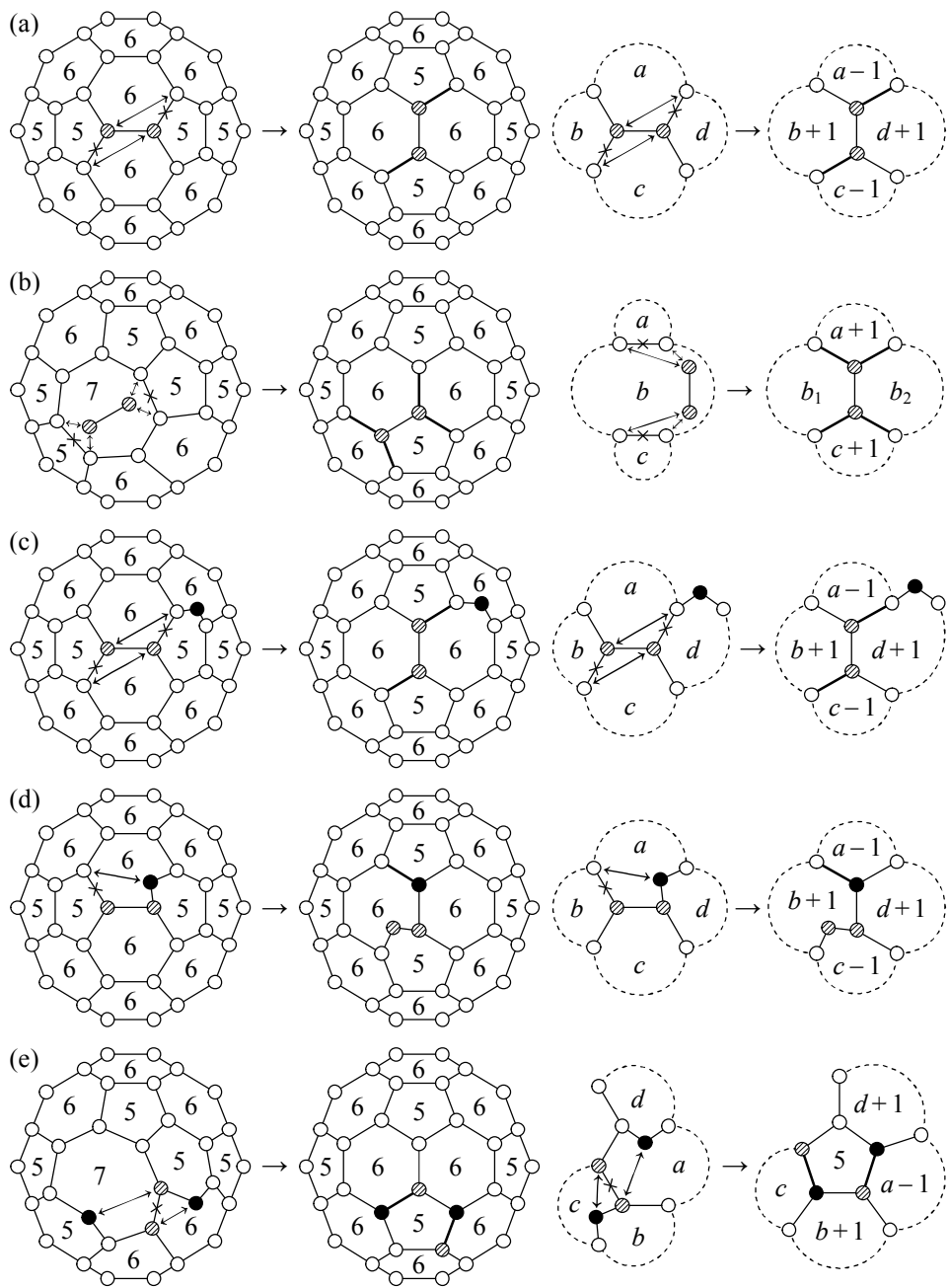

**Figure 2.** Schematic representation of atomic-scale reactions which change the numbers of atoms in polygons of the $sp^2$ structure of the fullerene shell. Left: (a), (b) and (e) reactions which lead to the $C_{60}$-$I_h$ fullerene, (c) and (d) reactions which lead to the $C_{61}$ fullerene with the same $sp^2$ structure as the $C_{60}$-$I_h$ fullerene but with an sp atom instead of one bond. Right: generalized schemes of these reactions. The double-headed arrow points to atoms that form a new bond after breaking the bond indicated by ×. The atom which was the sp atom before the reaction is colored in black. The new bond formed as a result of the reaction is shown in bold. Left: the numbers of

sp$^2$ atoms in polygons of the fullerene shell are indicated (the sp atom is not included in these numbers). Right: *a*, *b*, *c*, and *d* are the numbers of sp$^2$ atoms in polygons of the fullerene shell before the reaction (excluding the sp atom). (a) Stone–Wales (SW) reaction.[42] (b) Endo–Kroto mechanism of C$_2$ molecule insertion or emission.[71] Polygon *b* is divided into two parts, $b_1$ and $b_2$, it can be concluded that $(b_1 − 2) + (b_2 − 2) = b$. (c and d) SW reactions assisted by the sp atom: (c) the case where the same atom is sp before and after the reaction,[51] (d) the case where the sp atom becomes the sp$^2$ atom and vice versa one former sp$^2$ atom becomes the sp atom.[52] (f) Annihilation of a pair of sp atoms accompanied by pentagon formation.[2]

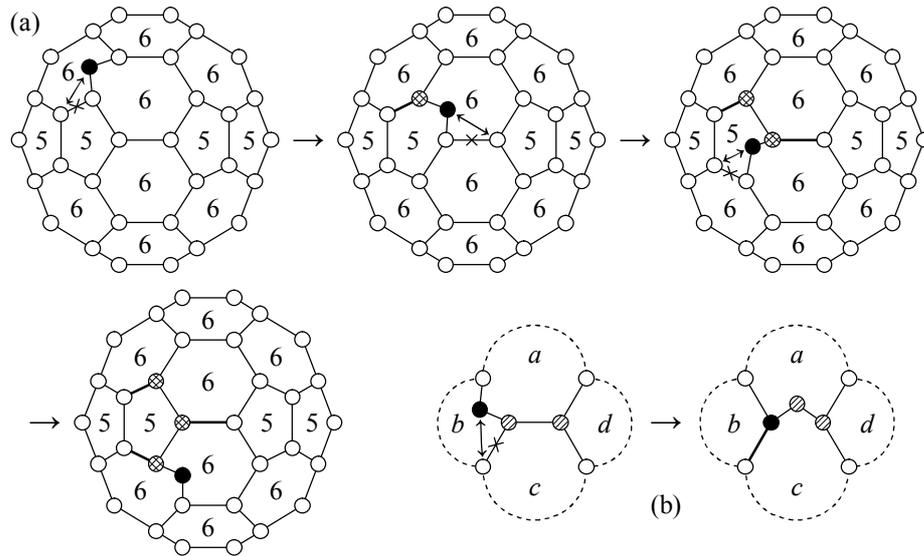

**Figure 3.** (a) Three subsequent sp-defect migration[2] events by the example of the C$_{61}$ fullerene with the same sp$^2$ structure as the C$_{60}$-$C_{2v}$ fullerene but with an sp atom instead of one bond. (b) Generalized scheme of sp-defect migration event. The double-headed arrows point to atoms that form a new bond after breaking the bond indicated by ×. The atom which was the sp atom before the reaction is colored in black. The new bond formed as a result of the reaction is shown in bold. (a) The numbers of sp$^2$ atoms in polygons of the fullerene shell are indicated (the sp atom is not included in these numbers); former sp atoms are crosshatched. (b) *a*, *b*, *c*, and *d* are the numbers of sp$^2$ atoms in polygons of the fullerene shell before the reaction (excluding the sp atom).

Originally, only atomic-scale reactions which preserve the sp$^2$ structure of the fullerene shell (that is only sp$^2$ atoms are present in the structure before and after the reaction) were proposed to explain selection of abundant fullerene isomers. Two types of such reactions, which keep the even number of atoms in a fullerene, were considered: bond rearrangements with a constant number of atoms and C$_2$ molecule emission/insertion. The simplest bond-rearrangement reaction in which the numbers of atoms in polygons of the fullerene sp$^2$ structure are changed is the Stone–Wales (SW) reaction.[42] The scheme of SW reaction which leads to formation of the C$_{60}$-$I_h$ isomer and the general scheme of SW reaction are shown in Figure 2a. All these reactions, the C$_2$ molecule emission,[24,43] C$_2$ molecule insertion,[32,44,45,46,47,48,49] and SW reaction[44,45,48,50] can in principle contribute into abundant fullerene isomer selection. However, later it was proposed that the main atomistic mechanism of selection of abundant fullerene isomers includes reactions in odd fullerenes with participation of sp atoms: bond-rearrangement reaction in which the numbers of atoms in polygons of the fullerene sp$^2$ structure are changed assisted by a nearby sp atom,[2,51,52,53,54] fullerene shell growth by insertion of single carbon atoms with subsequent annihilation of pairs of sp atoms,[2,55] insertion of a C$_2$ fragment from a C$_3$ molecule with formation of an odd fullerene at the intermediate stage,[45] insertion of a $C_2$ fragment from a longer carbon chain attached to a fullerene shell,[29] and emission of a $C_3$ molecule from the odd fullerene.[24]

Experimental results on atomistic mechanisms of selection of abundant fullerene isomers are scarce. Evidently, it is not possible to investigate experimentally intermediate structures of forming fullerene shells under the conditions when selection of abundant fullerene isomers takes place. Moreover, only the soluble part of the soot containing fullerenes can be analyzed[26], while the structure of the insoluble rest of soot is still unknown. We are not aware of any experimental studies of bond-rearrangements reactions which preserve the number of fullerene atoms. Nevertheless, some processes of fullerene growth and decrease of even fullerenes in size which can be indirectly related to the atomistic mechanisms of selection of abundant fullerene isomers listed above have been considered in the experiments. Based on the studies of distribution of kinetic energy release for accelerated fullerene ions produced by laser ablation of graphite, the energy required to emit a $C_2$ molecule was estimated to be 4.5, 4.6 and 3.5 eV for $C_{58}^+$, $C_{60}^+$ and $C_{62}^+$ ions, respectively.[43] However, the authors of this paper mentioned that the low values of these energies could be associated with highly exited states of the fullerene ions. The reaction rates for the $C_2$ molecule emission were measured in the experiments with laser ablation of polyimide[56] and graphite.[57] The lowest rates in comparison with other ions were detected for $C_{60}^+$ and $C_{70}^+$. The analysis of mass spectra of direct ions produced during laser ablation of fullerite containing $C_{60}$-$I_h$ and $C_{70}$-$D_{5h}$ showed that the $C_{60}^+$ ion is more stable relative to emission of the $C_2$ molecule than $C_{70}^+$.[58] This experiment also demonstrated the growth of the $C_{70}$ fullerene with formation of larger even fullerenes, while the growth of the $C_{60}$ fullerene did not take place under the same conditions. Note that reactions of $C_2$ molecule insertion and emission have not been observed directly and their possibility is a circumstantial conclusion based on the presence of even fullerene ions only in the mass spectra. The study of mass spectra of fullerenes formed during laser ablation of the $C_{60}$-$I_h$ isomer simultaneously with amorphous $^{13}C$ revealed generation of another abundant isomer, $C_{70}$-$D_{5h}$.[55] Based on the distribution of the number of $^{13}C$ atoms in the $C_{70}$-$D_{5h}$ isomer in these mass spectra, it was suggested that fullerene growth by single atom insertion and related bond-rearrangement reactions assisted by a nearby sp atom can provide a dominant contribution into atomistic mechanisms of selection of abundant fullerene isomers.

It should be emphasized that all these observations have been made under conditions that are different from the real conditions of fullerene synthesis (like arc discharge) when abundant isomers are formed. Therefore, these observations only demonstrate the possibility of such reactions and cannot prove that these reactions actually contribute into the atomistic mechanisms of selection of abundant fullerene isomers. Thus, theoretical modeling starting from consideration of the system at the atomic level holds the key to the determination of isomer selection mechanisms.

### 4. Modeling of fullerene formation

Although theoretical approaches should be crucial for resolving the mystery of formation of abundant fullerene isomers, there are a number of problems that complicate modeling of this process. In the present section we first discuss the inherent features of fullerene formation which lead to problems in atomistic modeling. Then we give an overview of different simulation methods used to study fullerene formation. After that results obtained by modeling on possible atomistic mechanisms of selection of abundant fullerene isomers are reviewed. And finally probabilities of different atomistic mechanisms to contribute notably into the isomer selection are considered.

*4.1 Problems in atomistic modeling*

The information on the atomistic mechanisms is normally obtained theoretically via atomistic modeling such as ab initio or classical MD. However, modeling of the whole fullerene formation process using only atomistic methods is problematic because of the following inherent features of the process.

First, fullerene formation is a multistage process. It can be divided into two main stages of formation of the initial fullerene shell via self-organization and later selection of abundant fullerene isomers by one or several of atomistic mechanisms listed above. Furthermore, it is not evident that formation of several abundant isomers from the initial mixture of numerous fullerene isomers of different sizes is a one-stage process. It is possible that several atomistic mechanisms with different time scales contribute comparably into the isomer selection. Note also that formation of initial fullerene shells due to self-organization of a chaotic carbon system is the process that is much faster than subsequent selection of abundant fullerene isomers (see, for example, Ref. [2,23]). Therefore, studies of fullerene shell formation and selection of abundant fullerene isomers need different simulation approaches. Here we consider mainly the strategy for modeling of abundant isomer selection.

Second, the number of atomic-scale reactions which take place between formation of the initial fullerene shell with numerous defects and final structure of a certain isomer is huge. For example, based on MD simulations,[32] it was concluded that all atoms of the fullerene can be replaced after the initial shell formation due to insertion and emission of $C_2$ molecules during selection of abundant isomers. The MD study of structural transformations in odd fullerenes[2] gave one further example of the large number of such reactions. Namely, it was found that only one of about 250 atomic-scale reactions occurs with a change in the polygons of the $sp^2$ structure and thus can in principle lead to selection of abundant fullerene isomers, while the majority of the reactions is migration of sp defect across the $sp^2$ structure, which remains the same.

And finally, a mixture of different isomers with a wide range of sizes should be formed initially via self-organization of carbon vapor. This statement was confirmed excellently by MD simulations where formation of fullerenes of different sizes and with numerous structural defects was observed under the same initial conditions of carbon vapor.[30,31]

Summarizing, the following inherent features of selection of abundant fullerenes make it difficult to study this process via atomistic simulations: 1) possibility that several atomistic mechanisms with different time scale simultaneously contribute into isomer selection; 2) a huge number of atomic-scale reactions on the way from the initial fullerene shell to a certain abundant isomer; 3) a variety of possible ways from a certain initial fullerene shell to a certain abundant isomer; 4) a great number of different isomers within a wide range of fullerene sizes in a mixture of initial fullerene shells.

*4.2 Simulation methods*

Let us now discuss advantages and disadvantages of different methods for modeling of fullerene formation. The set of methods used up to now to study fullerene formation includes MD simulations, quantum chemical calculations of activation barriers and energy changes for atomic-scale reaction related to fullerene formation, and kinetic models.

The main advantage of the reactive MD technique in fullerene formation studies is the possibility to directly observe non-equilibrium processes consisting of a huge number of atomic-scale reactions of bond rearrangements. As discussed above, formation of the initial fullerene shell with defects through self-organization has been revealed in this way for a wide set of starting carbon systems. Furthermore, MD simulations allow to discover new atomic-scale reactions that are not foreseen using imagination only. For example, the following reactions at last stage of initial fullerene shell formation have been revealed by MD simulations: detachment of carbon chains consisting of few atoms and attachment to the shell by one end[29] and insertion of such chains attached to the shell by both ends into the $sp^2$ structure of the shell.[23] As for the reactions taking place after the initial fullerene shell formation, the sp-defect migration in odd fullerenes,[2] emission of $C_3$ molecules from odd fullerenes,[24] transformation of an odd fullerene into an even fullerene via attachment of the $C_2$ molecule to an extra sp atom with subsequent emission of the $C_3$ molecule,[24] and the multistage open window mechanism of $C_2$ molecule emission from the $sp^2$ structure of the $C_{60}$-$I_h$ fullerene[59] have been discovered via MD simulations.

However, this advantage of MD simulations is useful only for relatively fast processes when the reaction probabilities (or reaction rates) are not too low in comparison with the frequencies of thermal vibrations of atoms near equilibrium positions. The maximal duration of the processes that can be modeled using this approach for carbon systems consisting of about 100 atoms is on the order of 1 μs for reactive classical MD simulations[2,23] and up to 5 ns for tight-binding MD simulations.[24] This time is considerably less than the formation time of the $C_{60}$ fullerene of 0.4 ms estimated for graphite laser ablation in the furnace filled with a buffer gas.[60] To deal with this problem, MD simulations related to fullerene formation are normally performed at an elevated temperature of 2400–4500 K for reactive classical MD simulations[2,23,61] and 2000–5600 K for tight-binding MD simulations.[29,30,30,31,32,25,59] The dependence of the reaction rate $p$ on the absolute temperature $T$ is determined by the Arrhenius equation as

$$p = \Omega \exp(-E_a/k_B T), \qquad (1)$$

where $\Omega$ is the pre-exponential factor, $E_a$ is the activation barrier of the reaction, $k_B$ is the Boltzmann constant. Thus, a high temperature allows to considerably decrease the computational time. It should be kept in mind, however, that any ratio of reaction rates also depends on temperature exponentially. Therefore, MD simulations at an elevated temperature allow to obtain only qualitative results for fullerene formation but the contributions of different atomic-scale reactions into the total process of selection of abundant fullerene isomers are not always realistic.

An additional issue one should take care of when performing MD simulations is related to the adequacy of reactive potentials for carbon. Significant efforts have been directed towards elaboration of potentials able to describe physical quantities important for fullerene formation.[62,63] The parameters of the last version REBO-1990EVC of first-generation Brenner potential were fitted to the energies of graphene edges, carbon chains, and monovacancy motion in a graphene layer, which makes this version adequate for simulations of the processes during the fullerene formation.[63] The adequacy of the REBO-1990EVC potential was confirmed by the good quantitative agreement of energy changes in sp-defect migration events in odd fullerenes with the values obtained by DFT calculations[2]. Nevertheless, only qualitative agreement was achieved for the activation barriers of such events.

DFT calculations allow to evaluate energy changes and activation energies of atomic-scale reactions. Using these data, the reaction probabilities can be estimated in accordance with the Arrhenius equation (1). Up to now, energy changes and activation barriers have been calculated using DFT-based methods for SW reactions in even fullerenes,[44,48,51,52,54,55,64,65] autocatalysis by an extra sp atom of SW reactions in odd fullerenes,[51,52,54,55] $C_2$ molecule insertion into even fullerenes,[44,47,48,54,55,66] and annihilation of pairs of sp atoms followed by formation of the $sp^2$ structure.[54] Energy changes for attachment of single carbon atoms have been also obtained.[54] Comparison of activation barriers calculated within the DFT framework and using empirical reactive potentials for reactions observed in MD simulations allows one to verify the MD results (see, e.g., the study[2] of sp-defect migration events in odd fullerenes).

The binding energy of fullerenes per one atom increases upon increasing the fullerene size and tends to the limit corresponding to graphite[44,49,67] (only $C_{60}$-$I_h$ and $C_{70}$-$D_{5h}$ have the binding energies slightly greater than $C_{62}$ and $C_{72}$ but lower than $C_{64}$ and $C_{74}$, respectively). Moreover, fullerene synthesis occurs under non-equilibrium conditions in the presence of energy and carbon sources. Thus, formation of abundant fullerene isomers has a kinetic origin. As discussed above, MD studies of selection of abundant fullerene isomers require extremely long simulations. This explains why development of kinetic models is important to solve the mystery of formation of $C_{60}$-$I_h$ and other abundant isomers. Note that such models should take into account atomic-scale reactions not only for insertion and emission of carbon atoms and small molecules by fullerenes but also for transitions between certain fullerene isomers. We are aware only one old study of fullerene formation using a kinetic model taking into account insertion and emission of small molecules without relation to the structure of certain fullerene isomers excluding only $C_{60}$-$I_h$ and $C_{70}$-$D_{5h}$.[45] For successful simulations of fullerene formation using kinetic models, rate constants for the reactions mentioned above are necessary. The rate constants for reactions of $C_2$ molecule insertion into the $sp^2$ structure of about ten fullerene isomers have been evaluated recently through DFT calculations of the activation barriers.[44]

None of the described methods of atomistic modeling is capable of dealing simultaneously with all four difficulties related to the inherent features of the process of abundant fullerene selection listed in the beginning of the section within a limited computing time. Different from atomistic modeling, kinetic models allow to overcome these difficulties and to consider a huge number of subsequent atomic-scale reactions even for a mixture of fullerene isomers within available computing time. To develop kinetic models able to consider atomistic mechanisms of abundant fullerene isomer selection, the structure of certain fullerene isomers should be taken into account. That is kinetic models should be supplemented by numerical description of fullerene isomers as well as of transitions between fullerene isomers via different atomic-scale reactions.

A method based on trivalent polyhedral graphs has been applied to enumerate all isomers of classical fullerenes[68] and even fullerenes with one tetragon[69] and one heptagon[69] in the structure. It has been found that a huge number of fullerene isomers are possible. For example, fullerenes $C_{60}$ and $C_{100}$ have 1812 and 570000 isomers, respectively, even if only pentagons and hexagons are present in their structure.[68] This number increases considerably if heptagons are also present.[69] Note that even fullerenes with tetragons and heptagons in their structure have been considered in atomistic modeling of fullerene isomer selection.[44,48,49,69] The number of atomic-scale reactions for transitions between fullerene isomers should be at least of the same order of magnitude. Polyhedral graphs have been used also to consider atomic-scale reactions of transformations of the $sp^2$ carbon network.[70] Thus, approaches based on polyhedral graphs allow to assign to each fullerene isomer a set of "adjacent" isomers related by different atomic-scale reactions. Such sets complemented by the calculated activation barriers and related reaction rates can be used in kinetic models of selection of abundant fullerene isomers.

### *4.3 Results of atomistic modeling*

Let us now briefly list the results related to selection of abundant fullerene isomers obtained by atomistic simulations for even and odd fullerenes. DFT calculations show that for majority of even fullerenes (except for the $C_{62}$ fullerene) there is a classical isomer (that is only with pentagons and hexagons in the structure) that has the lowest energy in comparison with nonclassical isomers with a single tetragon[69] or a single heptagon.[49,69] However, the energy difference between classical and nonclassical isomers with the lowest energies is roughly the same as the energy difference between classical isomers. Thus, participation of nonclassical isomers cannot be excluded even at the last stages of isomer selection. According to DFT calculations, SW reactions of bond rearrangements in even fullerenes have a huge activation barrier: 4.7 eV[51,52], 5.7 eV[54] for the transition from $C_{60}$-$C_{2v}$ to the $C_{60}$-$I_h$ fullerene, 6.2 eV[51], 7.0-7.3 eV[64] and 7.0-7.3 eV[65] for the back transition from $C_{60}$-$C_{2v}$ to the $C_{60}$-$I_h$ fullerene, 6.07 eV[48] for the transition between two isomers of the $C_{66}$ fullerene, 6.0 eV[44,48] for the transition from $C_{70}$-$C_s$ to the $C_{70}$-$D_{5h}$ fullerene, 7.5 eV[55] for the transition from $C_{70}$ with one heptagon in the $sp^2$ structure to the $C_{70}$-$D_{5h}$ fullerene. This makes these reactions unlikely under conditions of selection of abundant fullerene isomers (see, for example, calculations of the rate constant of the SW reaction[44]). Note that no reactions of bond rearrangements were observed during 0.5-1 μs at temperature 2500 K in classical MD simulations for eight different even fullerenes even when they had heptagons in the $sp^2$ structure or a one-coordinated atom attached.[23] Analogously, in other classical MD simulations at temperature 3000 K and during the total time 11.4 μs, such reactions were absent for different odd fullerenes in the part of shell with pure $sp^2$ structure, while about 17000 reactions of bond rearrangements with participation of an extra sp atom were detected during the same time.[2]

$C_2$ molecule insertion into the $sp^2$ structure of a fullerene shell occurs in two stages. First, a $C_2$ molecule attaches to the $sp^2$ structure without an activation barrier and then the attached molecule gets inserted into the $sp^2$ structure with an activation barrier.[44] The values of this barrier for different fullerenes range from 0.5 to 5 eV according to DFT calculations.[44,47,48,54,66] Barriers for a set of consecutive $C_2$ molecule insertion events starting from the $C_{50}$-$C_2$ fullerene and leading to formation of the $C_{60}$-$I_h$ fullerene and starting from the $C_{60}$-$I_h$ fullerene and leading to formation of the $C_{70}$-$C_s$ fullerene have been calculated.[44,48] Particularly, the barrier for $C_2$ molecule insertion into the $C_{58}$ fullerene with one heptagon in the sp$^2$ structure leading to formation of the $C_{60}$-$I_h$ fullerene was found to be 1.4 eV.[44] The computed values of the barrier indicate that the $C_2$ molecule insertion is possible under fullerene synthesis conditions. All these calculations of the insertion barrier, however, were performed for the simplest Endo–Kroto insertion mechanism[71] The scheme of Endo–Kroto $C_2$ insertion which leads to formation of the $C_{60}$-$I_h$ isomer and the general scheme of this reaction are shown in Figure 2b. Such a mechanism was not observed in tight-binding MD simulations,[32] while two different $C_2$ molecule insertion mechanisms were revealed in the same study.

According to the calculations, $C_2$ molecule emission from the sp$^2$ structure of a fullerene needs in total from 7.5 to about 14 eV for different fullerenes.[44,47,48,54,55,61,66] Particularly, 10.8 eV is necessary for $C_2$ molecule emission from the $C_{60}$-$I_h$ fullerene with formation of the $C_{58}$ fullerene with one heptagon in the sp$^2$ structure.[44] Such a huge energy change makes $C_2$ molecule emission unlikely at the last stage of selection of abundant fullerene isomers, even though the emission can proceed through intermediate metastable states with an even number of sp atoms. Tight-binding MD studies of $C_2$ molecule emission from the initial fullerene shell formed via self-organization of carbon vapor showed that *all* emission events occur at the defects of the sp$^2$ structure.[24,30] For even fullerenes, such defects are chains attached by both ends, pairs of an sp$^3$ atom and an sp atom and tetragons in the sp$^2$ structure[30], whereas $C_2$ molecules emitted from odd fullerenes include the former sp atom.[30,32] Emission of molecules different from $C_2$, up to $C_8$, from large fullerenes with numerous defects was observed in classical MD simulations at temperature 4100–4500 K.[61] However, no information on the mechanism of the emission events was provided in that paper. $C_2$ molecule emission also took place from the pure sp$^2$ structure of the $C_{60}$-$I_h$ fullerene in tight-binding MD simulations at very high temperature of 5600 K.[59]

As for bond rearrangements in odd fullerenes, DFT calculations show that the barriers for reactions promoted by an extra sp atom which are accompanied by changes in the number of atoms in polygons of the sp$^2$ structure are considerably lower than the barriers of SW reactions in the pure sp$^2$ structure mentioned above. Namely, the barrier of SW reactions assisted by an extra sp atom which lead to the $C_{61}$ fullerene with the same sp$^2$ structure as the $C_{60}$-$I_h$ fullerene but with an sp atom instead of one bond is 2.9 eV (and 4.0 eV is the barrier for the back reaction) in the case where the same atom is sp before and after the reaction [51] (see Figure 2c) and 1.1–1.3 eV[52,55] (and 2.3–2.5 eV[52,55] is the barrier for the back reaction) in the case where the sp atom becomes the sp$^2$ atom and vice versa one former sp$^2$ atom becomes the sp atom (see Figure 2d). The general schemes of these two reactions are also presented in Figure 2c,d. Other types of reactions that are assisted by an extra sp atom and lead to changes in the numbers of atoms in polygons of the sp$^2$ structure (analogously to SW reactions) are also possible, as follows from MD simulations.[2] According to DFT calculations, attachment of a single carbon atom to the sp$^2$ structure of a fullerene and its insertion into this structure in the place of a former bond with formation of an sp atom is a one-stage barrierless reaction[54] (different from two-stage $C_2$ molecule insertion described above). Therefore, it can be expected that reactions assisted by an extra sp atom in odd fullerenes should provide a dominant contribution to selection of abundant fullerene isomers in comparison with SW reactions and $C_2$ molecule insertion in even fullerenes.

Note that the same authors who considered the atomistic mechanism of formation of the $C_{60}$-$I_h$ and $C_{70}$-$D_{5h}$ isomers via $C_2$ molecule insertion,[44,47,48] investigated also the alternative mechanism with insertion of single carbon atoms.[54] Moreover, it was demonstrated by MD simulations that an extra sp atom in the $sp^2$ structure of an odd fullerene can easily move around the fullerene shell at temperature 3000 K (so-called sp-defect migration).[2] This bond-rearrangement reaction is analogous to the exchange mechanism for adatom migration on a surface in which is a new atom becomes the sp defect at each migration event. According to the DFT calculations by the example of 10 different sp-defect migration events in the $C_{69}$ fullerene, the activation barriers for such reactions lie in the range of 1.3–2.2 eV.[2] Because of the low values of the activation barriers for sp-defect migration, this can be the way in which the extra sp atom gets to the place of the fullerene shell where it assists to some reaction that contributes to formation of an abundant fullerene isomer. A set of sp-defect migration events is shown in Figure 3a by the example of the $C_{61}$ fullerene with the same $sp^2$ structure of the $C_{60}$-$C_{2v}$ fullerene but with an sp atom instead of one bond. The $sp^2$ structure of the $C_{60}$-$C_{2v}$ fullerene is chosen to show that sp-defect migration does not change the $sp^2$ structure and defects of the $sp^2$ structure remain the same (in this particular case, the presence of adjacent pentagons). The generalized scheme of sp-defect migration events is shown in Figure 3b.

Since all abundant fullerene isomers are even fullerenes, transformation of an odd fullerene to an even one is necessary after the reactions promoted by the extra sp atom. According to DFT calculations, emission of the extra sp atom from the $C_{69}$ and $C_{70}$ fullerenes needs in total 5–9 eV.[54] Such emission was not detected in MD simulations of an odd fullerene for the total time 11.4 μs at temperature 3000 K, while about 17000 reactions of bond rearrangements with participation of an extra sp atom were detected during the same time.[2] On the other hand, $C_2$ molecule emission from an odd fullerene was observed in classical and tight-binding MD simulations.[24,30,32,61] In the papers where the scheme of $C_2$ molecule emission from the odd fullerene is given, the $C_2$ molecule emitted includes the former sp atom.[30,32] Two ways of transformation of an odd fullerene to the even one via attachment of an additional carbon atom have been considered: (1) attachment of a carbon atom immediately to the extra sp atom with subsequent transformation of the formed defect to the pure $sp^2$ structure,[54] and (2) attachment of a carbon atom at an arbitrary place of the odd fullerene, subsequent migration of sp defects, and meeting and annihilation of the pair of sp atoms with formation of the $sp^2$ structure.[2,55] Such annihilation was observed in MD simulations in the $C_{66}$ fullerene.[2] The scheme of annihilation of the pair of sp atoms which leads to formation of the $C_{60}$-$I_h$ isomer and the general scheme of this reaction with formation of pentagon are shown in Figure 2e. Since attachment of an additional carbon atom to the arbitrary place of the odd fullerene is more probable than to the extra sp atom we assume that the second way is the main channel of this transformation. Attachment and insertion of short carbon chains[32] can also contribute to transformation of odd fullerenes to even ones.

In summary, based on the observations and results of atomistic modeling discussed above, one can consider two probable ways of selection of abundant fullerene isomers. The first way includes odd fullerenes as necessary intermediates between initial fullerene shells and specific abundant isomers, while in the second one, only even fullerenes are intermediates. The scheme comparing these two ways is presented in Table 1. The first way has advantages for selection of abundant isomers both for reactions which change fullerene size and reactions of bond rearrangements in which the fullerene size is maintained. Namely, insertion of a single carbon atom into the $sp^2$ structure of an even fullerene with formation of the sp defect is accompanied by sp-defect migration through the $sp^2$ structure. Therefore, reactions assisted by the sp atom which change the $sp^2$ structure and can lead to selection of abundant isomers can occur at any place which is necessary for such a selection and not related with the place of atom insertion. On the contrary, insertion of a $C_2$ molecule changes the $sp^2$ structure only at the place of $C_2$ molecule attachment. Thus, to obtain a specific isomer such an attachment should occur at a certain place of the fullerene shell. It is unlikely that $C_2$ molecule attachment at a certain place of the fullerene shell is more probable than at any other arbitrary place since this attachment is barrierless and occurs with a significant decrease of energy. As for reactions that change the $sp^2$ structure within the same fullerene size, such reactions for odd fullerenes are assisted by the sp atom and have a considerably lower barrier than the reactions for even fullerenes.

Table 1. Comparison of two possible ways of selection of abundant fullerene isomers: with both even and odd fullerenes (middle column) and with only even fullerenes (right column) as intermediates between initial fullerene shells and specific abundant isomers.

| Way of abundant fullerene isomer selection | with *odd and even* fullerenes as intermediates between arbitrary initial fullerene shell and specific abundant isomer | with *only even* fullerenes as intermediates between arbitrary initial fullerene shell and specific abundant isomer |
|---|---|---|
| Result of chaotic self-organization of a carbon system | mixture of odd and even fullerenes with defects | mixture of even fullerenes with defects |
| Reactions which change the fullerene size | insertion of single carbon atoms with subsequent migration of sp defects formed | insertion and emission of $C_2$ molecules |
| Reactions of bond rearrangements in which the fullerene size is maintained | changes of $sp^2$ structure promoted by sp atoms, and annihilation of pairs of sp atoms | SW reactions |
| Advantages/disadvantages for abundant fullerene isomer selection | **Advantages** 1) due to sp defect migration reactions assisted by the sp atom which change the $sp^2$ structure and can lead to selection of abundant isomers can occur *at any place* which is necessary for such a selection and not related with the place of atom insertion 2) *low* activation barriers for changes of $sp^2$ structure promoted by sp atoms lead to *high* reaction rates | **Disadvantages** 1) insertion of a $C_2$ molecule changes the $sp^2$ structure *only at the place* of $C_2$ molecule attachment 2) *high* activation barriers of SW reaction lead to *low* reaction rates |
| Conclusion | *highly probable way* of abundant fullerene isomers selection | *hardly probable way* of abundant fullerene isomers selection |

Therefore, we believe that the first way which includes odd fullerenes should provide the main contribution to selection of abundant fullerene isomers. For this way, the following four-stage atomistic mechanism has been proposed[2]: (1) attachment of single carbon atoms, (2) sp-defect migration to defects of the $sp^2$ structure such as heptagons or adjacent pentagons, (3) reactions assisted by the sp atom which lead to annealing of defects of the $sp^2$ structure and other changes of the numbers of atoms in polygons of the $sp^2$ structure and (4) annihilation of pairs of sp atoms. Reactions of fullerenes with $C_3$ and longer carbon chains which lead to transformation of odd fullerenes to even ones and vice versa can also be relevant and are consistent with the present four-stage mechanism. Nevertheless, the second way which occurs through emission and insertion of $C_2$ molecules into the $sp^2$ structure of even fullerenes as well as changes of the size of the even fullerene via attachment of carbon chains with subsequent detachment of longer or shorter chains cannot be excluded until the probabilities of these two ways are compared directly. The strategy of multiscale modeling (including atomistic approaches and kinetic models) which should make possible direct comparison of these two ways and determination of contributions of different atomistic mechanisms and certain reactions within this ways is considered in the next section.

## 5. Strategy of further multiscale modeling of fullerene formation

Let us discuss further studies that could help to solve the mystery of formation of abundant fullerene isomers based on different simulation approaches. For this purpose, we propose the following strategy divided into four research lines (see the scheme shown in Figure 4).

First, in accordance with the paradigm of fullerenes formation via self-organization, a mixture of even fullerene isomers (i.e. those with the pure $sp^2$ structure) and odd fullerene isomers (i.e. those with the $sp^2$ structure and a single sp atom) of different size should exist before selection of abundant fullerene isomers. The adequate choice of the initial set of classical and nonclassical fullerene isomers for further studies of selection of abundant fullerene isomers can be a difficult challenge. Reactions completing formation of the $sp^2$ structure of the fullerene shell (which determine the structure of initial fullerene shell) are still not well known. On the one hand, tight-binding MD simulations of fullerene shell formation in carbon vapor showed that in this case, the dominant contribution is provided by reactions which are accompanied by changes in the number of atoms in the forming fullerene shell (such as detachment of chains attached to the fullerene shell by one end and $C_2$ molecule insertion at defects of the $sp^2$ structure).[24,30,32] On the other hand, in classical MD simulation of fullerene shell formation as a result of transformation of an amorphous carbon cluster, reactions that conserve the number of atoms in the forming fullerene shell (such as insertion of chains attached to the fullerene shell by both ends into the $sp^2$ structure) were prevailing.[2] Classical MD allows to consider longer formation times and, therefore, lower temperatures that are closer to the experimental conditions. It can be useful to revise the simulations of formation of the initial fullerene shell in carbon vapor and determine schemes of various reactions completing formation of the $sp^2$ structure via classical MD simulations using the state-of-the-art reactive potential for carbon mentioned above.[63] The barriers of reactions found by MD simulations can be refined using *ab initio* methods to improve the description of the shell formation process.

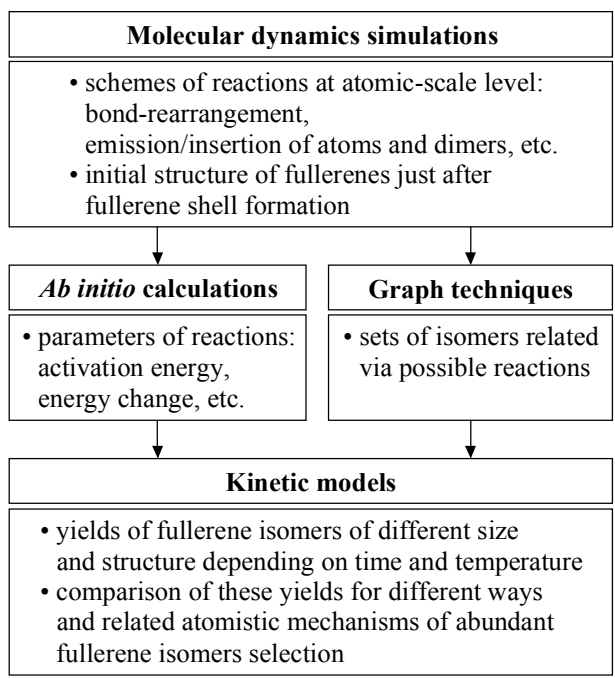

**Figure 4.** Scheme of multiscale atomistic simulations strategy for comparison of different pathways and related atomistic mechanisms of selection of abundant fullerene isomers.

Second, MD simulations can be used to reveal atomic-scale reactions related to selection of fullerene isomers which have not been foreseen using imagination only. In this way, reaction schemes can be obtained, for example, for reactions in odd fullerenes assisted by an extra sp atom which are accompanied by changes in the number of atoms in polygons of the $sp^2$ structure. An MD study of annihilation of sp atom pairs would be also useful for the same purpose. Among the tasks that can be studied using MD simulations, one can also investigate probabilities of attachment of carbon atoms, $C_2$ molecules and carbon chains to regions of the fullerene shell with a different local $sp^2$ structure as well as regions of the shell of an odd fullerene which contain the sp atom. It is also important to think on the schemes of subsequent insertion of attached $C_2$ molecules and chains into the $sp^2$ structure.

Third, the schemes of atomic-scale reactions revealed by MD simulations described above can be used to choose reactions for *ab initio* calculations of activation barriers and analyze the corresponding reaction probabilities. Such calculations need considerable computing power consumption. In this regard, the study of performance of a wide set of DFT, double-hybrid DFT and MP2-based methods in comparison with high-level *ab initio* methods for bond-reorganization reactions in eight $C_{60}$ isomers can be mentioned.[72] The reaction probabilities as functions of temperature can be estimated using the activation barriers obtained in the framework of the transition state theory.[44]

The fourth research line within the discussed strategy is application of kinetic models to study of selection of abundant fullerene isomers. We believe that sets of isomer pairs related by bond-rearrangement reactions for the same fullerene (such as SW reactions, changes of the $sp^2$ structure promoted by an sp atom, and annihilation of pairs of sp atoms) as well as isomer pairs related by insertion and emission of carbon atoms and $C_2$ molecules for different fullerenes can be obtained by numerical methods using graph theoretical techniques. Based on such sets of isomer pairs related by certain reactions and the values of the reaction probabilities dependent on temperature, the processes leading to selection of abundant fullerene isomers can be investigated by kinetic models under diverse experimental conditions and direct comparison of different pathways (or sets of subsequent reactions) from arbitrary initial fullerene shells with defects to specific abundant fullerene isomers can be performed. In this way, combination of the results of the four research lines within the proposed strategy of multiscale modeling will make it possible to solve the mystery of the high yield of abundant fullerene isomers.

**Funding**

AMP acknowledges the Russian Foundation of Basic Research (Grant No. 20-52-00035). IVL acknowledges the European Union MaX Center of Excellence (EU-H2020 Grant No. 824143). SAV and NAP acknowledge the Belarusian Republican Foundation for Fundamental Research (Grant No. F20R-301) and Belarusian National Research Program "Convergence-2025". Authors have no competing interests.

**References**

[1] Kroto, H.W.; Heath, J.R.; O'Brien, S.C.; Curl, R.F.; Smalley, R.E. $C_{60}$: Buckminsterfullerene. *Nature*, **1985**, *318*, 162–163.
[2] Sinitsa, A.S.; Lebedeva, I.V.; Polynskaya, Y.G.; Popov, A.M.; Knizhnik, A.A. Molecular Dynamics Study of sp-Defect Migration in Odd Fullerene: Possible Role in Synthesis of Abundant Isomers of Fullerenes. *J. Phys. Chem. C* **2020**, 124, 11652–11661.
[3] Kroto, H.W. The Stability of the Fullerenes $C_n$, with *n* = 24, 28, 32, 36, 50, 60 and 70. *Nature*, **1987**, *329*, 529–531.
[4] Poklonski, N.A.; Ratkevich, S.V.; Vyrko, S.A. Quantum-Chemical Calculation of Carbododecahedron Formation in Carbon Plasma. *J. Phys. Chem. A*, **2015**, *119*, 9133–9139.
[5] Lagow, R.J.; Kampa, J.J.; Wei, H.-C.; Battle, S.L.; Genge, J.W.; Laude, D.A.; Harper, C.J.; Bau, R.; Stevens, R.C.; Haw, J.F.; Munson, E. Synthesis of Linear Acetylenic Carbon: The "*sp*" Carbon Allotrope. *Science* **1995**, *267*, 362–367.
[6] Wakabayashi, T.; Shiromaru, H.; Kikuchi, K.; Achiba, Y. A Selective Isomer Growth of Fullerenes. *Chem. Phys. Lett.* **1993**, *201*, 470–474.
[7] Wakabayashi, T.; Achiba, Y. A Model for the $C_{60}$ and $C_{70}$ Growth Mechanism. *Chem. Phys. Lett.* **1992**, *190*, 465–468.
[8] Lebedeva, I.V.; Knizhnik, A.A.; Bagatur'yants, A.A.; Potapkin, B.V. Kinetics of 2D–3D Transformations of Carbon Nanostructures. *Physica E* **2008**, *40*, 2589–2595.
[9] Lozovik, Y.E.; Popov, A.M. Formation and Growth of Carbon Nanostructures: Fullerenes, Nanoparticles, Nanotubes and Cones. *Phys.-Usp.* **1997**, *40*, 717–737.
[10] Irle, S.; Zheng, G.S.; Wang, Z.; Morokuma, K. The $C_{60}$ Formation Puzzle "Solved": QM/MD Simulations Reveal the Shrinking Hot Giant Road of The Dynamic Fullerene Self-Assembly Mechanism. *J. Phys. Chem. B* **2006**, *110*, 14531–14545.
[11] Jones, R.O. Density Functional Study of Carbon Clusters $C_{2n}$ (2≤*n*≤16). I. Structure and Bonding in the Neutral Clusters. *J. Chem. Phys.* **1999**, *110*, 5189–5200.
[12] Killblane, C.; Gao, Y.; Shao, N.; Zeng, X.C. Search for Lowest-Energy Nonclassical Fullerenes III: $C_{22}$. *J. Phys. Chem. A* **2009**, *113*, 8839–8844.

[13] Portmann, S.; Galbraith, J.M.; Schaefer, H.F.; Scuseria, G.E.; Lüthi, H.P. Some New Structures of $C_{28}$. *Chem. Phys. Lett.* **1999**, *301*, 98–104.

[14] Hare, J.P.; Kroto, H.W.; Taylor, R. Preparation and UV/visible Spectra of Fullerenes $C_{60}$ and $C_{70}$. *Chem. Phys. Lett.* **1991**, *177*, 394–398.

[15] Rubin, Y.; Kahr, M.; Knobler, C.B.; Diederich, F.; Wilkins, C.L. The Higher Oxides of Carbon $C_{8n}O_{2n}$ (n = 3–5): Synthesis, Characterization, and X-Ray Crystal Structure. Formation of Cyclo[*n*]carbon Ions $C_n^+$ (*n* = 18, 24), $C_n^-$ (*n* = 18, 24, 30), and Higher Carbon Ions Including $C_{60}^+$ in Laser Desorption Fourier Transform Mass Spectrometric Experiments. *J. Am. Chem. Soc.* **1991**, *113*, 495–500.

[16] McElvany, S.W.; Ross, M.M.; Goroff, N.S.; Diederich, F. Cyclocarbon Coalescence: Mechanisms for Tailor-Made Fullerene Formation. *Science* **1993**, *259*, 1594–1596.

[17] Yeretzian, C.; Hansen, K.; Diederich, F.; Whetten, R.L. Coalescence Reactions of Fullerenes. *Nature* **1992**, *359*, 44–47.

[18] Chuvilin, A.; Kaiser, U.; Bichoutskaia, E.; Besley, N.A.; Khlobystov, A.N. Direct Transformation of Graphene to Fullerene. *Nat. Chem.* **2010**, *2*, 450–453.

[19] von Helden, G.; Gotts, N.G.; Bowers M.T. Experimental Evidence for the Formation of Fullerenes by Collisional Heating of Carbon Rings in the Gas Phase. *Nature* **1993**, *363*, 60–63.

[20] Hunter, J.; Fye, J.; Jarrold, M.F. Carbon Rings. *J. Phys. Chem.* **1993**, *97*, 3460–3462.

[21] Hunter, J.M.; Fye, J.L.; Jarrold, M.F. Annealing and Dissociation of Carbon Rings. *J. Chem. Phys.* **1993**, *99*, 1785–1795.

[22] Hunter, J.; Fye, J.; Jarrold, M.F. Annealing $C_{60}^+$: Synthesis of Fullerenes and Large Carbon Rings. *Science* **1993**, *260*, 784–786.

[23] Sinitsa, A.S.; Lebedeva, I.V.; Popov, A.M.; Knizhnik, A.A. Transformation of Amorphous Carbon Clusters to Fullerenes. *J. Phys. Chem. C* **2017**, *121*, 13396−13404.

[24] Irle, S.; Zheng, G.S.; Wang, Z.; Morokuma, K. Theory–Experiment Relationship of the "Shrinking Hot Giant" Road of Dynamic Fullerene Self-Assembly in Hot Carbon Vapor. *Nano* **2007**, *2*, 21–30.

[25] Irle, S.; Zheng, G.S.; Elstner, M.; Morokuma, K. From $C_2$ Molecules to Self-Assembled Fullerenes in Quantum Chemical Molecular Dynamics. *Nano Letters* **2003**, *3*, 1657–1664.

[26] Parker, D.H.; Wurz, P.; Chatterjee, K.; Lykke, K.R.; Hunt, J.E.; Pellin, M.J.; Hemminger, J.C.; Gruen, D.M.; Stock L.M. High-Yield Synthesis, Separation, and Mass-Spectrometric Characterization of Fullerenes $C_{60}$ to $C_{266}$. *J. Am. Chem. Soc.* **1991**, *113*, 7499–7503.

[27] Hahn, M.Y.; Honea, E.C.; Paguia, A.J.; Schriver, K.E.; Camarena, A.M.; Whetten, R.L. Magic Numbers in $C_N^+$ and $C_N^-$ Abundance Distributions. *Chem. Phys. Lett.* **1986**, *130*, 12–16.

[28] Bogana, M.; Ravagnan, L.; Casari, C.S.; Zivelonghi, A.; Baserga, A.; Li Bassi, A.; Bottani, C.E.; Vinati, S.; Salis, E.; Piseri, P.; Barborini, E.; Colombo, L.; Milani, P. Leaving the Fullerene Road: Presence and Stability of sp Chains in $sp^2$ Carbon Clusters and Cluster-Assembled Solids. *New J. Phys.* **2005**, *7*, 81.

[29] Zheng, G.S.; Irle, S.; Morokuma, K. Towards Formation of Buckminsterfullerene $C_{60}$ in Quantum Chemical Molecular Dynamics. *J. Chem. Phys.* **2005**, *122*, 014708

[30] Zheng, G.S.; Wang, Z.; Irle, S.; Morokuma, K. Quantum Chemical Molecular Dynamics Study of "Shrinking" of Hot Giant Fullerenes. *J. Nanosci. Nanotechnol.* **2007**, *7*, 1662–1669.

[31] Qian, H.-J.; Wang, Y.; Morokuma, K. Quantum Mechanical Simulation Reveals the Role of Cold Helium Atoms and the Coexistence of Bottom-Up and Top-Down Formation Mechanisms of Buckminsterfullerene from Carbon Vapor. *Carbon* **2017**, *114*, 635–641.

[32] Saha, B.; Irle, S.; Morokuma, K. Hot Giant Fullerenes Eject and Capture $C_2$ Molecules: QM/MD Simulations with Constant Density. *J. Phys. Chem. C* **2011**, *115*, 22707–22716.

[33] Chelikowsky, J.R. Formation of $C_{60}$ Clusters via Langevin Molecular Dynamics. *Phys. Rev. B* **1992**, *45*, 12062–12070.

[34] Makino, S.; Oda, T.; Hiwatari, Y. Classical Molecular Dynamics for the Formation Process of a Fullerene Molecule. *J. Phys. Chem. Solids* **1997**, *58*, 1845–1851.
[35] László, I. Formation of Cage-Like $C_{60}$ Clusters in Molecular-Dynamics Simulations. *Europhys. Lett.* **1998**, *44*, 741–746.
[36] Yamaguchi, Y.; Maruyama, S. A Molecular Dynamics Simulation of the Fullerene Formation Process. *Chem. Phys. Lett.* **1998**, *286*, 336–342.
[37] Yamaguchi, Y.; Maruyama, S. A Molecular Dynamics Study on the Formation of Metallofullerene. *Eur. Phys. J. D* **1999**, *9*, 385–388.
[38] Lebedeva, I.V.; Knizhnik, A.A.; Popov, A.M.; Potapkin, B.V. Ni-Assisted Transformation of Graphene Flakes to Fullerenes. *J. Phys. Chem. C* **2012**, *116*, 6572–6584.
[39] Irle, S.; Zheng, G.S.; Elstner, M.; Morokuma, K. Formation of Fullerene Molecules from Carbon Nanotubes: A Quantum Chemical Molecular Dynamics Study. *Nano Letters* **2003**, *3*, 465–470.
[40] Zheng, G.S.; Irle, S.; Elstner, M.; Morokuma, K. Quantum Chemical Molecular Dynamics Model Study of Fullerene Formation from Open-Ended Carbon Nanotubes. *J. Phys. Chem. A* **2004**, *108*, 3182–3194.
[41] Lee, G.-D.; Wang, C.Z.; Yu, J.; Yoon, E.; Ho, K.M. Heat-Induced Transformation of Nanodiamond into a Tube-Shaped Fullerene: A Molecular Dynamics Simulation. *Phys. Rev. Lett.* **2003**, *91*, 265701.
[42] Stone, A.J.; Wales, D.J. Theoretical Studies of Icosahedral $C_{60}$ and Some Related Species. *Chem. Phys. Lett.* **1986**, *128*, 501–503.
[43] Radi, P.P.; Hsu, M.-T.; Rincon, M.E.; Kemper, P.R.; Bowers, M.T. On the Structure, Reactivity and Relative Stability of the Large Carbon Cluster Ions $C_{62}^+$, $C_{60}^+$, and $C_{58}^+$. *Chem. Phys. Lett.* **1990**, *174*, 223–229.
[44] Li, M.-Y; Zhao, Y.-X; Han, Y.-B.; Yuan, K.; Nagase, S.; Ehara, M.; Zhao X. Theoretical Investigation of the Key Roles in Fullerene-Formation Mechanisms: Enantiomer and Enthalpy. *ACS Appl. Nano Mater.* **2020**, *3*, 547–554.
[45] Krestinin, A.V.; Moravsky, A.P. Mechanism of Fullerene Synthesis in the Arc Reactor. *Chem. Phys. Lett.* **1998**, *286*, 479–484.
[46] Saito, R.; Dresselhaus, G.; Dresselhaus, M.S. Topological Defects in Large Fullerenes. *Chem. Phys. Lett.* **1992**, *195*, 537–542.
[47] Dang, J.-S.; Wang, W.-W.; Zheng, J.-J.; Zhao, X.; Osawa, E.; Nagase S. Fullerene Genetic Code: Inheritable Stability and Regioselective $C_2$ Assembly. *J. Phys. Chem. C* **2012**, *116*, 16233–16239.
[48] Wang, W.-W.; Dang, J.-S.; Zheng, J.-J.; Zhao, X.; Nagase S. Selective Growth of Fullerenes from $C_{60}$ to $C_{70}$: Inherent Geometrical Connectivity Hidden in Discrete Experimental Evidence. *J. Phys. Chem. C* **2013**, *117*, 2349–2357.
[49] Hernández, E.; Ordejón, P.; Terrones, H. Fullerene Growth and the Role of Nonclassical Isomers. *Phys. Rev. B* **2001**, *63*, 193403.
[50] Walsh, T.R.; Wales, D.J. Relaxation Dynamics of $C_{60}$. *J. Chem. Phys.* **1998**, *109*, 6691–6700.
[51] Eggen, B.R.; Heggie, M.I.; Jungnickel, G.; Latham, C.D.; Jones, R.; Briddon, P.R. Autocatalysis During Fullerene Growth. *Science* **1996**, *272*, 87–89.
[52] Ewels, C.P.; Heggie, M.I.; Briddon, P.R. Adatoms and Nanoengineering of Carbon. *Chem. Phys. Lett.* **2002**, *351*, 178–182.
[53] Lee, I.-H.; Jun, S.; Kim, H.; Kim, S.Y.; Lee, Y. Adatom-Assisted Structural Transformations of Fullerenes. *Appl. Phys. Lett.* **2006**, *88*, 011913.
[54] Wang, W.-W.; Dang, J.-S.; Zhao, X.; Nagase S. Revisit of the Saito–Dresselhaus–Dresselhaus $C_2$ Ingestion Model: on the Mechanism of Atomic-Carbon-Participated Fullerene Growth. *Nanoscale* **2017**, *9*, 16742–16748.

[55] Dunk, P.W.; Kaiser, N.K.; Hendrickson, C.L.; Quinn, J.P.; Ewels, C.P.; Nakanishi, Y.; Sasaki, Y.; Shinohara, H.; Marshall, A.G.; Kroto, H.W. Closed Network Growth of Fullerenes. *Nat. Comm.* **2012**, *3*, 855.

[56] Campbell, E.E.B.; Ulmer, G.; Busmann, H.-G.; Hertel I.V. Stability and Fragmentation of Carbon Clusters. *Chem. Phys. Lett.* **1990**, *175*, 505–510.

[57] Radi, P.P; Hsu, M.T.; Brodbelt-Lustig, J.; Rincon, M.; Bowers M.T. Evaporation of Covalent Clusters: Unimolecular Decay of Energized Size-Selected Carbon Cluster Ions ($C_n^+$, $5 \leq n \leq 100$). *J. Chem. Phys.* **1990**, *92*, 4817–4822.

[58] Ulmer, G.; Campbell, E.E.B.; Kühnle R.; Busmann, H.-G.; Hertel I.V. Laser Mass Spectroscopic Investigations of Purified, Laboratory-Produced $C_{60}/C_{70}$. *Chem. Phys. Lett.* **1991**, *182*, 114–119.

[59] Xu, C.; Scuseria, G.E. Tight-Binding Molecular Dynamics Simulations of Fullerene Annealing and Fragmentation. *Phys. Rev. Lett.* **1994**, *72*, 669–672.

[60] Suzuki, S.; Yamaguchi, H.; Ishigaki, T.; Sen, R.; Kataura, H.; Krätschmer, W.; Achiba, Y. Time Evolution of Emission by Carbon Nanoparticles Generated with a Laser Furnace Technique. *Eur. Phys. J. D* **2001**, *16*, 369−372.

[61] Srinivasan, S.G.; van Duin, A.C.T.; Ganesh, P. Development of a ReaxFF Potential for Carbon Condensed Phases and Its Application to the Thermal Fragmentation of a Large Fullerene. *J. Phys. Chem. A* **2015**, 1*19*, 571–580.

[62] Sinitsa, A.S.; Lebedeva, I.V.; Knizhnik, A.A.; Popov, A.M.; Skowron, S.T.; Bichoutskaia, E. Formation of Nickel–Carbon Heterofullerenes under Electron Irradiation. *Dalton Trans.* **2014**, *43*, 7499–7513.

[63] Sinitsa, A.S.; Lebedeva, I.V.; Popov, A.M.; Knizhnik, A.A. Long Triple Carbon Chains Formation by Heat Treatment of Graphene Nanoribbon: Molecular Dynamics Study with Revised Brenner Potential. *Carbon* **2018**, *140*, 543–556.

[64] Bettinger, H.F.; Yakobson, B.I.; Scuseria, G.E. Scratching the Surface of Buckminsterfullerene: The Barriers for Stone–Wales Transformation through Symmetric and Asymmetric Transition States. *J. Am. Chem. Soc.* **2003**, *125*, 5572–5580.

[65] Kumeda, Y.; Wales, D.J. Ab Initio Study of Rearrangements between $C_{60}$ Fullerenes. *Chem. Phys. Lett.* **2003**, *374*, 125–131.

[66] Jin, C.; Lan, H.; Suenaga, K.; Peng, L.; Iijima, S. Metal Atom Catalyzed Enlargement of Fullerenes. *Phys. Rev. Lett.* **2008**, *101*, 176102.

[67] Curl, R.F.; Lee, M.K.; Scuseria, G.E. $C_{60}$ Buckminsterfullerene High Yields Unraveled. *J. Phys. Chem. A* **2008**, *112*, 11951–11955.

[68] Fowler, P.W.; Manolopoulos, D.E. *An Atlas of Fullerenes*; Clarendon Press: Oxford, 1995.

[69] Gan, L.-H.; Wu, R.; Tian, J.-L.; Clarke, J.; Gibson, C.; Fowler, P.W. From $C_{58}$ to $C_{62}$ and Back: Stability, Structural Similarity, and Ring Current. *J. Comp. Chem.* **2017**, *38*, 144–151.

[70] Ori, O.; Cataldo, F. Moving Pentagons on Nanocones. *Fullerenes, Nanotubes and Carbon Nanostructures* **2020**, *28*, 732–736.

[71] Endo, M.; Kroto, H.W. Formation of Carbon Nanofibers. *J. Phys. Chem.* **1992**, *96*, 6941–6944.

[72] Karton, A.; Waite, S.L.; Page A.J. Performance of DFT for $C_{60}$ Isomerization Energies: A Noticeable Exception to Jacob's Ladder. *J. Phys. Chem. A* **2019**, *123*, 257–266.